\documentclass[prl,showpacs, twocolumn,groupedaddress]{revtex4}

\usepackage{graphicx}

\begin{document}

\title
{Metallic behaviour of carrier-polarized 
C$_{60}$ molecular layers: Experiment and Theory
}

\author
{
Z.H. Lu, C. C. Lo, C. J. Huang}
\affiliation{Department of Materials Science and Engineering,
 University of Toronto,
184 College Street, Toronto, Ontario M5S 3E4
}
\author
{ M.W.C. Dharma-wardana
}
\email[Email address:\ ]{chandre.dharma-wardana@nrc.ca}
\author
{Marek Z. Zgierski}
\affiliation{
National Research Council of Canada, Ottawa, Canada. K1A 0R6\\
}

%
%
%
\date{\today}
\begin{abstract}
Although C$_{60}$ is a molecular crystal with a bandgap $E_g$ of $\sim 2.5$ eV,
we show that $E_g$ is strongly affected by
injected charge.
In sharp contrast to the Coulomb blockade typical of
quantum dots,  $E_g$ is {\it reduced} by the
Coulomb effects.
The conductance of a thin C$_{60}$ layer sandwiched 
between  metal (Al, Ag, Au, Mg and Pt) contacts is investigated.
Excellent Ohmic conductance 
is observed for Al electrodes protected with ultra-thin
LiF layers. First-principles 
calculations, Hubbard models etc.,
show that the 
energy gap of C$_{60}$ is dramatically reduced
when electrons hop from C$_{60}^-$ to C$_{60}$.
\end{abstract}
%
\maketitle
%
The Fullerene C$_{60}$ solid is a molecular
crystal\cite{sawatsky}                  
 with a bandgap of $\sim 2.5$ eV. 
Alkali-metal doping makes
it a conductor or a
superconductor\cite{gunnerson}.    
%
 Other methods of converting C$_{60}$
into a conductor involve exotic chemical routes where, e.g., 
 C$_{60}$ molecules
are joined by metal-ligand structures\cite{jacs}.
Such efforts focus on improving the
conductance of the molecular layer itself.
Another important factor is the 
electrical contact between the molecular layer and the
metal surface.
The practical realization of molecular electronics  depends
crucially on this molecule/metal contact. In many situations,
such contacts
 involve chemical processes that lead to fragmentation
 of the 
 molecules themselves\cite{critique,turak}. 

Fullerenes, e.g., C$_{60}$ are popular 
 candidates for molecular electronics. Unlike  quantum dots
and related nanostructures, molecules come in identical copies and have
 energy
levels which are robust at room temperature. 
 Many studies on moleculer-C$_{60}$
 devices have  emphasized
negative-differential resistance (NDR)\cite{ndr},
or the control of conductance using scanning-tunnelling probes,
 electromechanical
 and other gates\cite{guo}.                      
In the simulation by Taylor  et al\cite{guo},  a single C$_{60}$ molecule is
  positioned
between two Al leads. In their ideal device,
metallic conductance occurs when the triad of 
degenerate LUMO (lowest-unoccupied molecular orbital) acquires
 three electrons
forming a half-filled band. Hubbard-like effects 
are assumed  negligible in their calculations; the LUMO is $\sim$ 1.8 eV
above the HOMO (highest occupied molecular orbital) in the
local-density approximation (LDA).

Given that   single-molecule current ($I$) 
and voltage ($V$) data  are
still  controversial\cite{critique},
we turn to the 
the $I-V$ data
 for  C$_{60}$ films sandwiched 
between two  metal electrodes.
Usually the top electrode is formed by  depositing hot metal
onto the organic film already deposited on 
a cold, crystallographically uncontrolled metal surface.

Studies of C$_{60}$ layers deposited on crystallographically 
controlled surfaces of Al, Ag, Au, Mg,
 etc.\cite{c60al,auag}, show that       
charge transfer, electronic and
bonding modifications
occur within the first monolayer of molecules,
 but the second monolayer usually
remains unaffected\cite{auc60layer1,agbell},   
except for weaker physical effects.
 More than one C$_{60}$ monolayer may be affected if the surface is
polycrystalline.
We call this physico-chemically  modified
interface-layer the {\it metal-fulleride} layer.
The metal-fulleride formation for C$_{60}$ deposition on Al,
as well as for Al deposition on C$_{60}$ has been studied
by, e.g., Owens et al.\cite{c60al}.    
 The interaction is more
complex than resulting from image-charge effects or simple charge transfer,
as may perhaps be the case for Ag. 
The situation is even more complicated for Pt and Ni surfaces
which form strong bonds with C$_{60}$ and catalyse
the decomposition of the organic molecule at sufficiently
elevated temperatures\cite{PtNi}.      
If the ``bulk'' C$_{60}$ material
is clean and undoped, very few carriers are available for producing
``band-bending'' effects of  dipole layers etc., as found at doped
inorganic semi-conductor interfaces.
 This implies that, except for the
fulleride layer next to the electrode,
the molecules in the ``bulk'' are unaffected
by the presence of the metal contacts. This picture assumes that
hot-metal atoms were not
bombarded into the organic film\cite{c60al,agbell},
 or that the metal was not presented
as a paste or other preparation
 where metal diffusion into the C$_{60}$
could  occur\cite{ohmic1}.    
%
%
%

Most reports\cite{katz,sarici} of    
$I-V$ behaviour on such {\it metal}/C$_{60}$/{\it metal}
sandwich structures
 suggest semiconducting, rectifying or
insulating behaviour.
 In a previous paper\cite{apl1}      
 we showed how
 an ultra-thin layer of
LiF dramatically modifies the $I-V$ characteristics of a
 molecular film by (i) protecting
the organic molecules from reactions with the metal,
 (ii) creating a sharp electrode
density of states (DOS) and favouring carrier injection.
Similar important effects of ionic epilayers on metals are
not uncommon\cite{ag_mgo}.                                   
 Here we study
 C$_{60}$ molecular films and show that the charge injection
dramatically changes the ``insulating'' or semi-conducting character of 
C$_{60}$ films. Metallic conductance is achieved for M = Al and Mg. That is,
the current $I$ depends linearly on the applied bias $V$, for both
forward and reverse bias.
 When M=Pt, the conductance is very small,
 non-linear  and there
 is strong asymmetry on reversing the bias. We note that the
work functions for polycrystalline
 Mg, Ag, Al, Au and Pt are
3.7, 4.3, 4.3, 5.1 and 5.7 eV
 respectively\cite{workfunc}.      
 Although
there is some  correlation with the work function,
the situation is more complex since carrier hopping is subject to
 Coulomb interactions as well.

Even if the C$_{60}$ monolayer (fulleride layer) just next to the metal electrode
acquires electrons due to interactions with the metal\cite{agbell},
conduction cannot occur unless there
 is a mechanism for further  charge transfer
from molecule to molecule\cite{hopping}    
 in undoped C$_{60}$.  
We clarify the observed  metallic behaviour
using  detailed electronic structure calculations. 
The energy gap of the ``bulk'' 
C$_{60}$ molecules dynamically collapses as the carrier electrons hop from
molecule to molecule. That is, an inverse Coulomb blockade occurs, where
charge transfer between
the highly polarizable C$_{60}$ molecule is {\it enhanced}.
 In metals like Pt where fulleride formation via metal $d$-orbitals
is likely,
 charge injection is negligible  under low bias;
here strongly non-linear $I-V$ is observed.
 Thus while neutral
C$_{60}$ behaves as {\it a molecular crystal in photoemission}
experiments\cite{sawatsky}, our experiments and calculations
suggest that 
 C$_{60}$ may behave as {\it a metal in the presence of polarizing
carriers}.

{\it Experimental.--}
Details of sample fabrication etc.,
are given in Lu et al.\cite{apl1}.
The devices were made on 2~in$\times$2~in Si(100)
 wafers with 2000 nm furnace oxides on top.
 The first metal (Al, Mg, Ag, Au, Pt) electrode
 (1 mm wide, 50 mm long, and 60 nm thick)
 was deposited through a shadow mask, and is referred to as
 the ``bottom'' electrode. C$_{60}$ films
 (210 nm thick) were then deposited over the bottom electrode.
 A second metal (Al) 
electrode (1~mm wide, 50mm long and 100 nm thick), referred to
 as the ``top electrode'',
 was deposited over the C$_{60}$ films. An ultra-thin layer
 ($\sim$ 1 nm) of LiF 
was deposited on the C$_{60}$ film prior to the deposition of the
top electrode. The top electrode lines were orthogonal to the
 bottom electrode lines so that each intersection of these
 two lines produces one
{\it metal}/C$_{60}$/{\it metal} device.
 There are 20 to 100 devices on each wafer.
 A final silicon
 oxide film of $\sim$ 300 nm was used to encapsulate the device.
 This final encapsulation
 is essential to produce consistent and reproducible results.
Such protection is important since contamination with
air has a drastic effect\cite{air}  on the electrical properties of
C$_{60}$.
 All devices
 were made using a K.J. Lesker  4~in~ x~ 4~in  Cluster Tool having several process
 chambers inter-linked through a central distribution chamber. All metals were
 deposited in the metallization
chamber having a base pressure of $\sim$10$^{-7}$ Torr. The C$_{60}$ molecules
were deposited in the organic chamber having a base pressure of $\sim10^{-8}$ Torr. The
transfer of samples between  chambers
 was via a central chamber, with a base pressure of $\sim10^{-9}$ Torr.
 The $I-V$ measurements were
 in a dark, ambient environment,
 using an HP4140B meter with a Materials Development probe
station.

\begin {figure}
\includegraphics*[width=7.0cm, height=8.0cm]{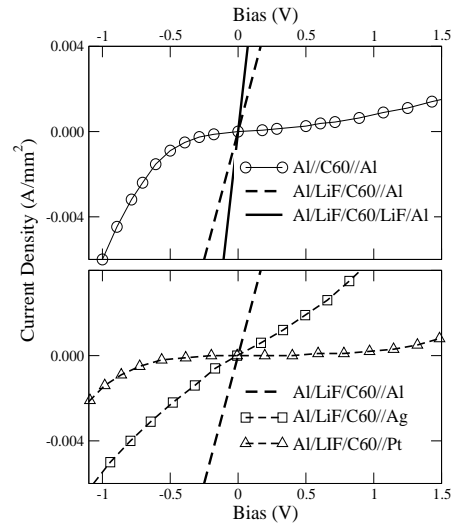}
\caption
{Top:
Top panel: Experimental $I-V$ curves
for three Al/X/C$_{60}$/X/Al devices, with X=LiF
or left unprotected (//) at the
top and bottom interlayers.
Bottom panel: $I-V$ curves for Al/LiF/C$_{60}$//M, with 
 M=Al, Ag and Pt. Note that LiF is present only at the top
Al electrode.
}
\label{oneIV}
\end{figure}
%
%

 Figure.~1,
top panel
 shows the $I-V$ data for
%
samples which differ in the presence or absence of LiF interlayers.
In samples with bare-Al electrodes , i.e., Al//C$_{60}$//Al
the left-hand Al (the ``top'' electrode) was deposited hot on C$_{60}$,
while the right-hand Al electrode (``bottom'' electrode) was cold during
C$_{60}$ deposition. In such systems the 
fulleride interlayers
are uncontrolled and they are denoted by, //, a double bar.
In devices with LiF at the top electrode,
i.e,  Al/LiF/C$_{60}$//Al, the C$_{60}$ is
protected from the hot deposition of the ``top'' Al electrode.

The device with bare Al electrodes produces a typical
diode-like $I-V$
behaviour (see Fig.~1, top panel)
with negligible current flow at low bias. The
reverse-bias current flow ``takes off'' more rapidly, as seen in the
figure.
 The use of a LiF
interlayer at the top electrode (the ``hot-deposited'' electrode) is enough to
create a linear $I-V$ relationship with zero threshold
 for current flow. This is typical
of metallic conductors. When
a LiF interlayer is included at the bottom electrode as well, the conductance 
($I/V$) improves only slightly, showing that the bottom electrode, where
C$_{60}$ was deposited in the cold, required no LiF protection.
 The behaviour is completely metallic, 
with no discontinuity in the gradient $I/V$ on changing the sign of $V$.
This was confirmed for C$_{60}$ layer varying
 from 100 nm to 250 nm in thickness.
It is clear that Al$^{3+}$ migration into the C$_{60}$ layer cannot be the
cause of the conductance (as was the case in Ref.~\cite{ohmic1}),
since  such migration is  obstructed by the LiF layer. 
Photoemission experiments suggest that only the first few monolayers
of C$_{60}$ near the metal are affected by the presence of the metal\cite{auag}.
The $I-V$ data show that fulleride formation at the cold-deposited
C$_{60}$ on the Al interface is not critical to electron transport,
while the prevention of fulleride formation at the hot-deposited
Al on C$_{60}$ is absolutely important. The improved
conductance of the sample with the Al electrodes protected with LiF
confirms the protective role of LiF. This also suggests that the
resistance of the device is determined by that of the fullaride layer
which plays the role of the least conducting link.

Figure 1, bottom panel shows typical results
for Al, Ag and Pt. Both 
Al and Mg (not shown) show metallic behaviour. However, surprisingly,
Ag  shows some slight  non-linearity in its $I-V$, revealing
some interface reactivity\cite{auag}. Results for Au (not shown)
tend to be significantly more non-linear than for Ag. The
Pt/C$_{60}$ interface  is clearly very unfavourable for Ohmic $I-V$.

\begin{figure}
\includegraphics*[width=8.0cm, height=9.0cm]{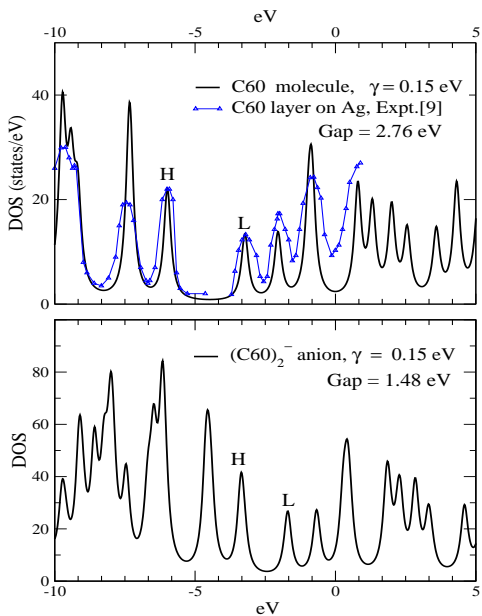}
\caption
{(a)The ectron density of states (DOS) for isolated C$_{60}$
near the HOMO-LUMO gap. Experimental results\cite{auag} for C$_{60}$ on
Ag are shown for comparison.(b) DOS for the C$_{60}-$C$_{60}^-$ anion,
 the presence of the charge polarizes the system, lowering
the gap.
See Table I.
}
\label{figdos}
\end{figure}

{\it Theory.--}
We have studied the HOMO-LUMO energy gap $E_g$
 as well as the
electronic density of states (DOS)
for the C$_{60}$ molecule, for C$_{60}-$LiF, C$_{60}-$C$_{60}$,
C$_{60}-$C$_{60}^-$ and C$_{60}-$C$_{60}^+$ structures, where the
short bar $-$ stands for all the interactions between the two moities.
These electronic-structure details are obtained from
density functional calculations using
 the Gaussian-98 code\cite{G98}        
 at the
BP86/6-31G*
level\cite{acro,G98}.     
 The calculations included geometry
optimization via total energy minimization using
gradient corrected exchange-correlation functionals. The modifications
in the DOS of some of the  C$_{60}$ structures
are shown in Fig.~\ref{figdos}, and in table I.
In the upper panel we compare our isolated-molecule DOS with
the experimental results\cite{auag} for multilayers of C$_{60}$ on Ag obtained
using direct and inverse photoemission (UPS and IPS). Thus the HOMO part of the
experimental curve is from UPS (energy resolution $\sim 0.1$ eV), 
and the LUMO part of the curve is from
 IPS (energy resolution $\sim 0.3$ eV). We
have positioned these experimental curves so that our
 calculated HOMO (H) and LUMO (L)
peaks align with the experimental H and L peaks. The general
 agreement clearly confirms
that the molecules in a C$_{60}$ multilayer deposited on Ag 
 are essentially like isolated molecules. This also serves to confirm the
validity of our theoretical calculations.

The presence of  LiF itself produces a 
 decrease in the  C$_{60}$
energy gap.
Here the distance between the LiF and the C$_{60}$
center, as well as other inter-nuclear distances have been 
energy optimized, and hence this is the upper-bound to the
gap reduction that may arise from LiF.

The energetics of the C$_{60}$ interactions is
 given in Table~\ref{energies}.
The energy gaps $E_{g}$ given in the Table improve on 
the gaps calculated using the local-density approximation (LDA).
Thus the LDA gap for C$_{60}$ is only  $\sim 1.8$ eV, 
and differs from the experimental
  solid-state gap of $\sim 2.5$ eV.
Our spin density-functional calculations show a spin-polarized
energy-level structure for the anion.
The electrons of the C$_{60}$ system form a low-density
quasi-2D electron fluid. Such systems have spin dependent
ground states under suitable conditions\cite{pd2d}. This may be of
importance in spin transport applications using
suitable C$_{60}$ based systems\cite{pasupathy}, although here the 
spin splitting is only about $\sim 0.02$ eV.
\begin{table}
\caption{The HOMO, LUMO energies 
$E_L$ and $E_H$, for the
C$_{60}$ system are given below. 
 The LiF is symmetrically placed orthogonal to a
hexagonal ring of C$_{60}$.
Th C$_{60}-C_{60}$  inter-center distance for the
anion or the cation is
10.2 \AA. The ``LUMO'' of the anion is occupied by an electron,
while the ''HOMO'' of the cation is occupied by a hole.
The energy gaps are from the fully optimized BP86/6-31G*
and not from LDA. The $\alpha$, $\beta$ refer to
spin states of the electron or hole.
}
\begin{ruledtabular}
\begin{tabular}{ccccc}
system     & $E_H$ a.u. & $E_L$ a.u. & $E_{g}$ eV. \\
\hline \\
%
C60       &  -0.22003 & -0.11854 & 2.8   \\
%
%
%
C$_{60}-$LiF    & -0.18993 & -0.16157 & 0.8 \\
%
%
%
%
%
%
C$_{60}-$C$_{60}^-\alpha$&-0.11880&-0.06443&1.48 \\
C$_{60}-$C$_{60}^-\beta$ &-0.11736&-0.06213&1.50 \\
%
%
C$_{60}-$C$_{60}^+\alpha$&-0.28683&-0.23093&1.52 \\
C$_{60}-$C$_{60}^+\beta$ &-0.28485 &-0.22942&1.51 \\
%
%
\label{energies}
\end{tabular}
\end{ruledtabular}
\end{table}
%
 
Photoemission experiments\cite{ c60al,auag} clearly establish
the metallization of the first $C_{60}$
 layer on metals like Ag and Al,
with as much
 as 1.8 electrons transferred\cite{sawat4}     
to the LUMO in the case of Ag. That this is close to 2 electrons
per molecule suggests that the Hubbard like on-site repulsion
has been reduced from that of the isolated molecule.
 An electron transferred to a C$_{60}$ molecule, and
subject to the applied bias,
hops to an adjacent C$_{60}$ when carrier transport occurs.
 The final state of such a
hop is given in our calculations for the 
(C$_{60}-$C$_{60}$)$^-$ anion, and shows a reduced gap.
This is due to (a) splitting of the five-fold HOMO
and the three-fold LUMO multiplets by the electric field
of the
in-coming charge,
 (b) polarization and distortion
 of the of the molecules which persist
under stationary state conditions,
 (c) resulting
modification of the on-site
 Coulomb interactions. Many of these
issues have been examined
 using Hubbard type models\cite{sawatPol},     
or with microscopic approaches. Hesper et al.\cite{hesper}
used an image-charge model for C$_{60}$ on Ag to obtain
a gap reduction of $\sim 1.44$ eV.  They even mention
the possibility of "driving the insulator into the metallic state".
Our explicite calculations (Table I)
can be used to parametrize the Hubbard models.
 However, even the Hubbard-model 
conductance  has
not been evaluated,
 except in special cases (e.g, infinite-dimensional
Hubbard models). 

In the usual picture 
of electron transport across molecular devices like:
{\it  A/Molecular-layer/B},
 it is assumed that an electron is injected
from the source electrode {\it A}
 into the LUMO of the  nearest molecule,
 converting it to a transient, excited  $M^{-*}$ anion.
This involves Coulomb blockade, 
rearrangement of bond lengths,
bond angles etc.,  to give the actual  $M^-$ anion. 
At this point the carrier may become localized on the molecular
site as a poloron. Then
no conduction  occurs until a suitably high bias is applied.
Or, if the energy offsets are favourable, the carriers may hop to
neighbouring  molecules and successively move towards the drain electrode.
In the devices discussed here, the fulleride layers  for Al, Mg
and Ag are 
already populated at the LUMO with up to $\sim$ 1.8 electrons. 
Thus the 1st layer of C$_{60}$, i.e., the metal-fulleride layer contains
charge which
polarizes the second layer and
enables the hopping of electrons under the applied
bias. The usual $C_{60}$ van der Waals crystal
now behaves as a conductor. The conductance is determined by the
weakest factor, viz., the 
transmission coefficient $T_{fb}$ for the process, 
metal-fulleride $\to$ C$_{60}$ bulk-like layer.
 An evaluation 
of $T_{fb}$ is postponed till a detailed fulleride-C$_{60}$
 calculation becomes available.

Our geometry optimizied calculations (Table I) 
 includes the electron-phonon interaction
to all orders. The importance of  these effects is recognized,
especially within the alkali-doped fullerides like K$_4$C$_{60}$ which
is an insulator due to bond-distortion effects. Also, K$_6$C$_{60}$ is
a band insulator since the 6 electrons per molecule 
completely fill the three-fold LUMO. In our system, the
LUMO of the fulleride layer is only occupied
 to $\le$ 1.8 electrons\cite{sawat4} per C$_{60}$ molecule. 

 In conclusion, we have shown, experimentally and theoretically, that
 $I-V$ characteristics 
similar to a metal can be obtained
 using LiF protected low-workfunction electrodes connected
to C$_{60}$ layers.

{\it acknowledgements-}  We thank George Sawatzky (UBC) for
 his valuable comments.

\end{document}